\documentclass[runningheads]{llncs}
\usepackage{graphicx}
\usepackage{amsmath}
\usepackage{amssymb}
\usepackage[misc]{ifsym}
\usepackage{makecell}
\usepackage{listings}
\usepackage{CJKutf8}
\usepackage{xcolor}
\usepackage{listings}
\usepackage{multirow}
\begin{document}
\begin{CJK}{UTF8}{gbsn}
\title{ASGNet: Adaptive Semantic Gate Networks for Log-Based Anomaly Diagnosis 
}
%
%
%
%
%

%

\author{Haitian Yang\inst{1,2}{(\Letter)}\and
Degang Sun\inst{2}{(\Letter)}\and
Wen Liu \inst{1}\and\\
Yanshu Li\inst{1,2}\and
Yan Wang \inst{1,2}\and
Weiqing Huang\inst{1,2}}
\authorrunning{Yang et al.}
%
\institute{ Institute of Information Engineering,Chinese Academy of Sciences,Beijing,China \\
\email{$\{$yanghaitian$\}$@iie.ac.cn}\\ \and
School of Cyber Security,University of Chinese Academy of Sciences,Beijing,China
}

\maketitle              
\begin{abstract}

Logs are widely used in the development and maintenance of software systems. Logs can help engineers understand the runtime behavior of systems and diagnose system failures. For anomaly diagnosis, existing methods generally use log event data extracted from historical logs to build diagnostic models. However, we find that existing methods do not make full use of two types of features, (1) statistical features: some inherent statistical features in log data, such as word frequency and abnormal label distribution, are not well exploited. Compared with log raw data, statistical features are deterministic and naturally compatible with corresponding tasks. (2) semantic features: Logs contain the execution logic behind software systems, thus log statements share deep semantic relationships. How to effectively combine statistical features and semantic features in log data to improve the performance of log anomaly diagnosis is the key point of this paper. In this paper, we propose an adaptive semantic gate networks (ASGNet) that combines statistical features and semantic features to selectively use statistical features to consolidate log text semantic representation. Specifically, ASGNet encodes statistical features via a variational encoding module and fuses useful information through a well-designed adaptive semantic threshold mechanism. The threshold mechanism introduces the information flow into the classifier based on the confidence of the semantic features in the decision, which is conducive to training a robust classifier and can solve the overfitting problem caused by the use of statistical features. The experimental results on the real data set show that our method proposed is superior to all baseline methods in terms of various performance indicators.

\end{abstract}

\section{Introduction}
\begin{figure}[htbp]
\centering
\includegraphics[scale=1]{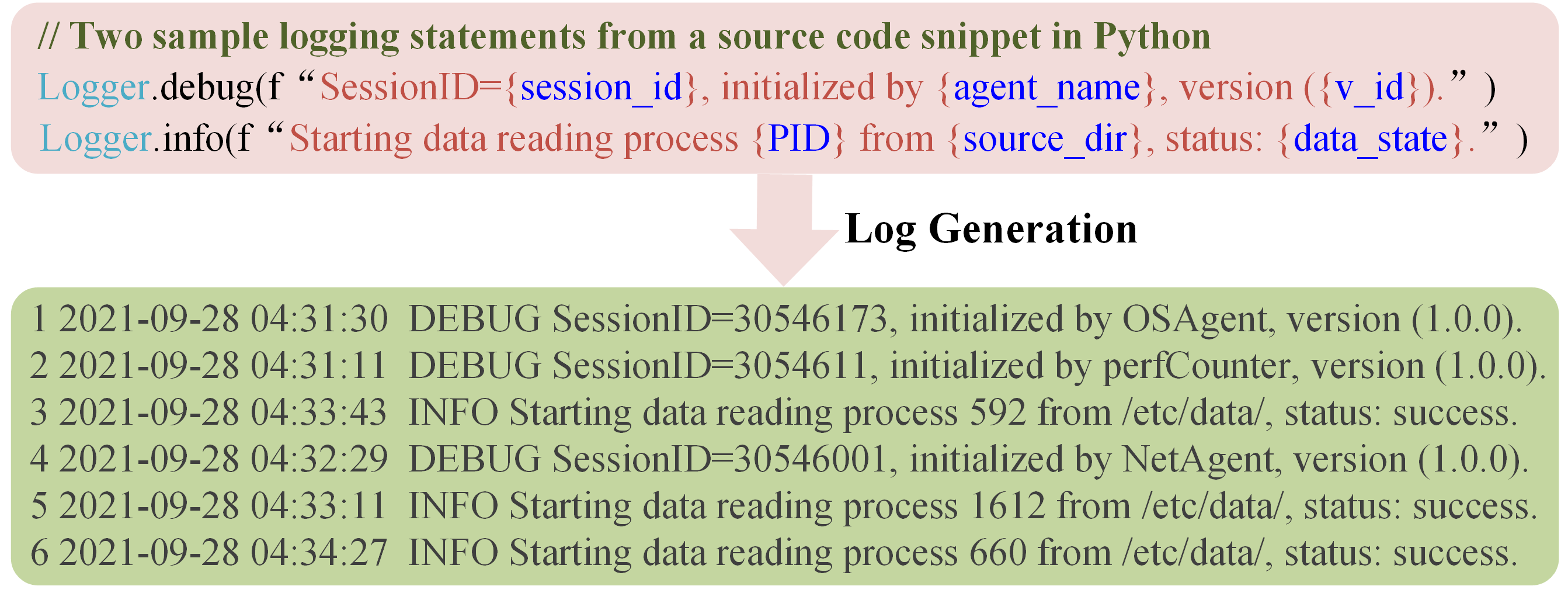}
\caption{The process from initialization to sending data.}
\label{f1}
\end{figure}

With the rapid development and evolution of information technology, During the past few years, we witness that large-scale distributed systems and cloud computing systems gradually become critical technical support of the IT industry \cite{1}. Anomaly detection and diagnosis play an important role in the event management of large-scale systems\cite{23,24}, which aims to detect abnormal behavior of the system in time. Timely anomaly detection enables system developers (or engineers) to pinpoint problems the first time and resolve them immediately, thereby reducing system downtime \cite{2,3,6}. However, as the scale of modern software become larger and more complex, the traditional log anomaly detection and diagnosis approaches based on specialized domain knowledge or manually constructed and maintained rules become less and less inefficient \cite{7,8,9,19}. Benefiting from the development of deep learning technology, a number of effective log anomaly detection and diagnosis methods emerge in recent years, but these methods ignore two issues\cite{20,22}, (1) Statistical features: some inherent statistical features in log data, such as word frequency and abnormality label distribution, is not well utilized by deep learning based methods. Statistical features\cite{30,31} consist of statistical characteristics deterministic compared with log raw data, which is naturally compatible with the corresponding task. (2) Semantic features: Logs contain the execution logic behind the software system, thus log statements share deep semantic relationships. Hence, this paper focuses on how to effectively combine statistical features and semantic features in log data to improve the performance of log anomaly diagnosis. In order to demonstrate the problems that we raised in log data, we list different examples in Table \ref{tab1} and Figure \ref{f1}.

\begin{table*}[htbp]

    \centering
    \caption{Statistics on the number of occurrences of words in the seven log datasets}
    \label{tab:my_label}
    \resizebox{\textwidth}{!}{
    \begin{tabular}{cccccccc}
    \hline
         Dataset & BGL & BGP & Tbird & Spirit & HDFS & Liberty & Zokeeper \\
    \hline
         Dataset size &1.207GB &1.04GB &27.367GB &30.289GB &1.58GB &29.5GB &10.4M\\
     \hline
        total number of distinct words &5632912 &4491076 &23330854 &12793353 &3585666 &14682493 &53094\\ 
    \hline
        \multirow{2}*{appear only once} &5173492 &4330501 &7789598 &3861462 &2576220 &2020068 &28954 \\
         & (91.8\%) &(96.4\%) &(33.4\%) &(30.2\%) &(71.8\%) &(13.8\%) &(54.5\%)\\
    \hline
        \multirow{2}*{appear less than 5 times} &5298053 &4424922 &16863767 &9914926 &2853030 &6656253 &51443 \\
         & (94.05\%) &(98.5\%) &(72.3\%) &(77.5\%) &(79.6\%) &(45.3\%) &(96.9\%)\\
    \hline
        \multirow{2}*{appear less than 10 times} &5403556 &4463325 &19948834 &10815055 &2885414 &9882010 &52686 \\
         &(95.9\%) &(99.4\%) &(85.5\%) &(84.5\%) &(80.5\%) &(67.3\%) &(99.23\%)\\
    \hline
        \multirow{2}*{appear less than 20 times} &5465876 &4472566 &21112093 &11311509 &3353119 &11425515 &52797\\
         &(97.03\%) &(99.6\%) &(90.5\%) &(88.4\%) &(93.5\%) &(77.8\%) &(99.4\%)\\
    \hline
        \multirow{2}*{appear at least once per 10000 lines} &3825 &1299 &79063 &137224 &1998 &10052 &564\\
         &(0.068\%) &(0.029\%) &(0.34\%) &(1.07\%) &(0.056\%) &(0.068\%) &(1.06\%)\\
    \hline
        \multirow{2}*{appear at least once per 1000 lines} &573 &316 &6243 &2663 &258 &5365 &172\\
         &(0.01\%) &(0.007\%) &(0.027\%) &(0.021\%) &(0.007\%) &(0.037\%) &(0.32\%)\\
    \hline
    \end{tabular}
}    
    \label{tab1}
\end{table*}

From Table \ref{tab1}, we can see that most of the words are infrequent, and most of these infrequent words appear only once. For example, in the BGP dataset, 96.4\% of words appear only one time. In the Liberty dataset, 77.8\% of the words have a frequency lower than 20. In the BGP and Zookeeper datasets, the number of words with a frequency lower than 20 accounts for more than 99\%. At the same time, only a small fraction of words are frequent, i.e. they occur at least once in every 10000 or 1000 lines of logs. We can observe that most of the logs generated during the system operation are normal, and the abnormal logs are limited. Based on this, we infer that the abnormal words are not frequent, further we conclude that statistical features in logs are necessary for anomaly diagnosis.

From Figure \ref{f1}, we can see that the first log to the sixth log shows the complete process from initialization to sending data, demonstrating the complete program execution logic. Therefore, log sequence contains the execution logic behind the program and has rich semantic information. We can see from Table \ref{tab1} and Figure \ref{f1}, that the statistical features and text semantic features used in this paper are widely present in log data.

To deal with the two above-mentioned issues, this paper designs an Adaptive Semantic Gate Networks (ASGNet) which consists of three parts, log statistics information representation(V-Net),log deep demantic representation(S-Net) and adaptive semantic threshold mechanism(G-Net). Specifically, V-Net leverages an unsupervised autoencoder\cite{11} to learn a global representation of each statistical feature vector, where we note that employing variational inference can further improve model performance compared to vanilla autoencoders. S-Net extracts latent semantic representations from text input by pre-trained RoBERTa\cite{10}. G-Net aligns information from two sources, then adjusts the information flow. 

The main contributions of this paper are summarized as follows:

(1) We attempt to explicitly leverage statistical features of system log data for anomaly diagnosis in a deep learning architecture. We also demonstrate that our proposed approach is very effective based on various experiments.

(2) To fuse statistical features into low-confidence semantic features, we propose a novel adaptive semantic threshold mechanism to retrieve necessary and useful global information. The experimental results prove that the threshold mechanism is very effective for the log-based anomaly diagnosis task.

(3) We conduct extensive experiments on 7 datasets of different scales and subjects. Results show that our proposed model yields a significant improvement over the baseline models. 

\section{Related Work}
The main purpose of anomaly diagnosis is to help O$\&$M engineers analyse the cause of anomalies and understand the operational status of the system or network. A number of excellent methods have emerged in recent years.

Yu et al.\cite{25} utilized log templates to build workflows offline. Such models can provide contextual log messages of problems and diagnose problems buried deep within log sequences, which can provide the correct log sequence and tell engineers what is happening. Jia et al.\cite{26} proposed a black-box diagnostic method TCFG (Timed-weight Control Flow Graph) for control flow graphs with temporal weights, which does not require prior domain knowledge and any assumptions and achieves better performance on relevant datasets.

Fu et al.\cite{27} proposed the use of a Finite Automata Machine (FSA) to simulate the execution behaviour of a log-based system model. The model first clusters logs to generate log templates and removes all parameters based on regular expressions. Each log is then labelled by its log template type to construct sequences from which the FSA is learned to capture the system's normal work flow, achieving better performance on relevant datasets. Beschastnikh et al.\cite{28} generated finite state machines from the execution trace of a concurrent system to infer a concise and accurate model of that system's behaviour. Engineers can use the inferred finite state machine model of communication to understand complex behaviour and detect anomalies for developers. Lou et al.\cite{29} proposed an automaton model and a corresponding mining algorithm for reconstructing concurrent workflows. The algorithm can automatically discover program workflows and can construct concurrent workflows based on traces of interleaved events.

Although these methods have achieved better performance, none of these explored statistical feature and semantic feature of log sequence. Our study breaks the conventional thinking of treating logs as general objects of time series and introduces novel ideas in natural language processing to anomaly detection and diagnosis, investigating log sequences as text sequences with semantic information. 

\section{The Proposed Model}

\begin{figure}[htbp]
\begin{center}
\includegraphics[width=\textwidth]{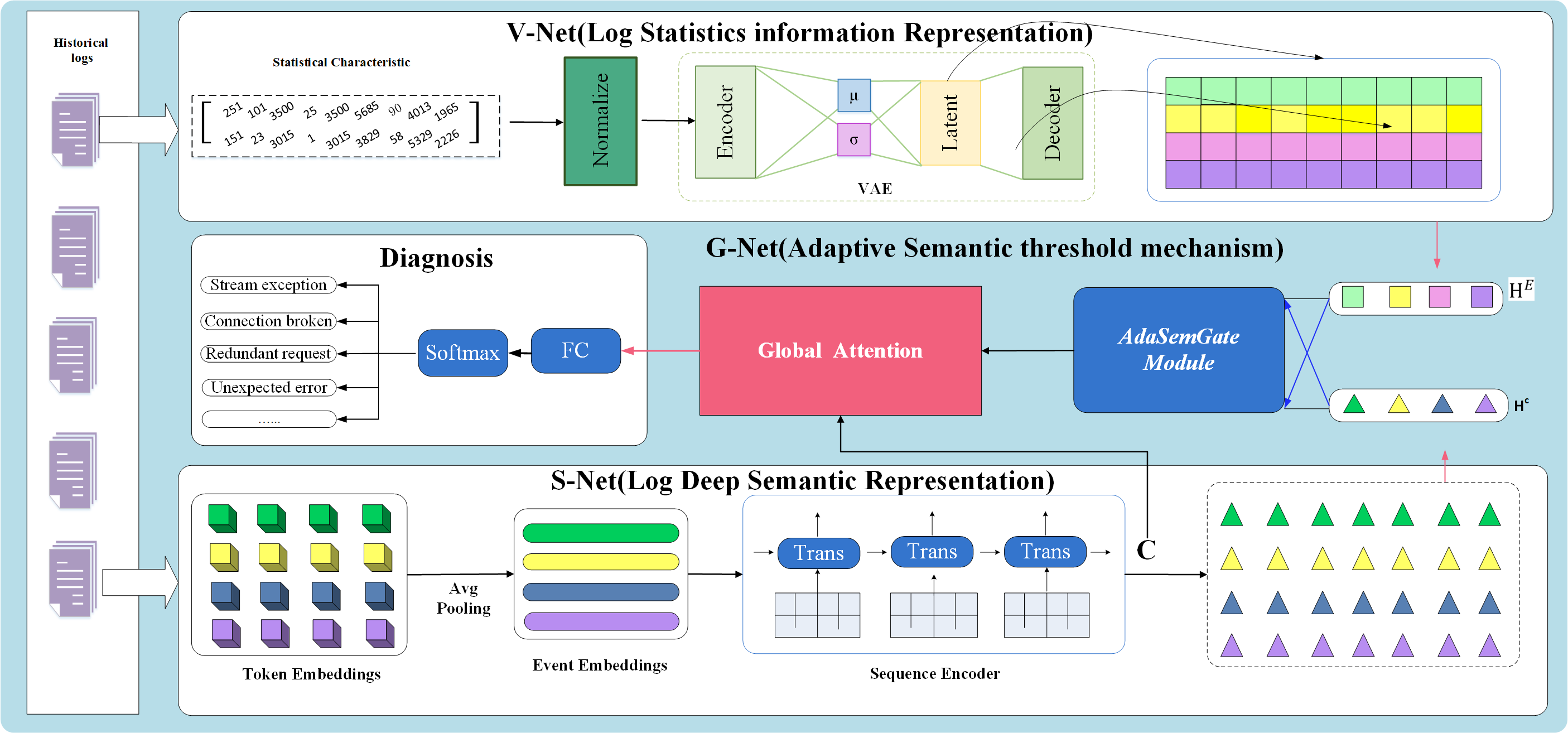}
\end{center}
\caption{Overview of our proposed ASGNet model.}
\label{f2}
\end{figure}

In this section, we present our proposed log-based anomaly diagnosis ASGNet model. Model structure is depicted in Figure \ref{f2}.

\subsection{Task description}
In this research, the log-based anomaly diagnosis task can be described as a tuple of three elements $(S,I,y)$, where $S=[s^1,s^2,\dots,s^g ]$ represents the log sequence  whose length is $g$. $I$ denotes the current log message task ID and $y\in Y$ conveys the anomaly diagnosis specific labels for logging exceptions, currently diagnosable exceptions are Stream exception, Connection broken, Redundant request, Unexpected error, etc. 

\subsection{Definition of terms}

In this section, we first give the formal definition of the log statistics vector, which is defined as follows. Given a word $w$ in a log message and a set of log exception labels $C=\{c_1,c_2,\cdots,c_n,\}$, the statistical information vector of $w$ is:

 \begin{equation}
     E^w=[e_1,e_1,\cdots,e_n]
 \end{equation}

\noindent where $e_i$ represents the count of word $w$ on label $c_i$.
Given a log message $L=\{w_i\}_{i=1}^m$, the word statistics matrix in the log message is as follows:

\begin{equation}
    E^L=[E^{w_1},E^{w_2},…,E^{w_m}]
\end{equation}

The statistical information vector captures the global distribution of log anomaly labels as a feature of the words in the log. 

These statistical features are primitive but can be used for feature selection by determining word-relatedness. Intuitively, if a word $w$ has very high or very low frequency across all labels, then we can assume that $w$ has a limited contribution to the anomaly diagnosis task. In contrast, if a word appears more frequently in specific label class, we assume this word is discriminative. Here, the term statistical label vector dictionary $E$ is only obtained from the training set.

\subsection{Log Statistics information Representation}
Inspired by the text classification \cite{30},Log Statistics information Representation (V-Net) aims to convert statistical features into an efficient representation. Log statistics vectors contain integer counts of words, thus initially incompatible with semantic features in both dimension and scale. Therefore, V-Net employs an autoencoder to map discrete vectors of log statistics into a latent continuous space to obtain a global representation of the statistics. In this work, we employ a variational autoencoder(VAE)\cite{11} to encode log statistics vectors.

We generate log statistics vectors for all log messages in the data set to obtain $S=\{E_{(i)}^L\}_{i=1}^N$, which consists of $N$ discrete log statistics Vector composition. We assume that all log statistics vectors are generated by a stochastic process $p\theta(E|S)$ involving a latent variable S sampled from a prior distribution $p\theta(S)$. Since the posterior $p\theta(S|E)$ is intractable, we cannot directly learn the generative model parameters $\theta$. Next, we employ a variational approximation $q\phi(S|E)$ to jointly learn the variational parameters 	$\phi$ and $\theta$. Therefore, we can optimize the model by maximizing the marginal likelihood that is composed of a sum over the marginal likelihoods of individual $E$:

\begin{equation}
    log⁡p_{\theta}(E)=D_{KL}(q_{\phi}(S|E)|| p_{\theta} (S∣E))+	{\L}(\theta,\phi;E)
\end{equation}

Because the $KL$ divergence term is non-negative, we can derive the likelihood term ${\L}(\theta,\phi; E)$ to obtain a variational lower bound on the marginal likelihood, i.e.:

\begin{equation}
    {\L}(\theta,\phi;E) =-D_{KL} (q_{\phi} (S∣E)|| p_{\theta} (S)) +E_(q_{\phi}(S|E)) [log⁡p_{\theta} (E∣S)] )
\end{equation}

\noindent where the $KL$ term is the closed solution, and the expected term is the reconstruction error. We employ a reparameterization approach to adapt the variational framework to an autoencoder.
We use two encoders to generate two sets of $\mu$ and $\sigma$ as the mean and standard deviation of the prior distributions, respectively. Since our approximate prior is a multivariate Gaussian distribution, we represent the variational posterior with a diagonal covariance structure:

\begin{equation}
    logq_{\phi}(S|E) =log⁡N(S;\mu,\sigma^2 I)
\end{equation}

By training an unsupervised VAE model, we can obtain latent variables $E^s$ through a probabilistic encoder, which would be a global representation of statistical features. The training of V-Net is independent of the main classifier, and the representation $E^s$ is generated in the preprocessing stage and is fed to the classifier through the adaptive semantic threshold mechanism.

\subsection{Log Deep Semantic Representation}

Log Deep Semantic Representation (S-Net) extracts semantic features from log message input and projects the semantic features into the information space for confidence evaluation. The input of S-Net is a log message $W=[w_i,w_2,\cdots,w_m,]$ with fixed length $m$.

In this paper, we use the pre-training model to obtain the semantic representation of the log. Specifically, we use the pre-trained RoBERTa\cite{10} to extract the feature map of the input log text:

\begin{equation}
    C=RoBERTa(W)
\end{equation}

Then, we map the semantic feature map $C$ into the information space through dense layers. We use the sigmoid to activate values in the representation.

\begin{equation}
    H^C=W^C\cdot C+b^C
\end{equation}

\noindent where $H^{'C}= \sigma（H^C）$, where $\sigma(\cdot)$ is the sigmoid function, are used to evaluate the confidence of the corresponding semantic features in the decision-making process.

\subsection{Adaptive Semantic threshold mechanism}
As described in Section 2.3, the semantic representation of log statistics features $E^s$ is obtained offline. To flexibly utilize statistical features, we apply dense layers to project $E^s$ into the information space shared with semantic features:

\begin{equation}
    H^E=W^E\odot (E^s)+b^E
\end{equation}

The valve component is fused with $H^C$ and $H^E$, and the semantic feature map $H^O$ enhanced by statistical information is output through the AdaSemGate function,

\begin{equation}
    H^O = AdaSemGate⁡(H^C,H^{'C},H^E,\epsilon ) \\
    = ReLU⁡(H^C )+Gate⁡(H^{'C},\epsilon)\odot H^E 
\end{equation}

\noindent where $ReLU(·)$ is the activation function, and ⊙ represents element-wise point multiplication. The values in $H^{'C}$ are in probabilistic form, and the Gate function is designed to recover less confident entries (probability close to 0.5) to match elements in $H\zeta$. Specifically, for each unit $a\in H^{'C}$,

\begin{equation}
    Gate⁡(\alpha,\epsilon) =\begin{cases} \alpha,  & ``if " 0.5-\epsilon \ge\alpha \le 0.5+\epsilon   \\ 0, & ``otherwise " \end{cases}
\end{equation}

\noindent where $\epsilon$ is a vulnerable hyperparameter for tuning the confidence threshold. Specifically, if $\epsilon = 0$, all statistics are rejected, and if $\epsilon=0.5$, statistics are accepted. Therefore, the $Gate⁡(\alpha,\epsilon)$ function uses element-wise multiplication as a filter to extract only the necessary information.

We employ global attention to combine the consolidated semantic representation $H^O$ with the original feature map $C$:

\begin{equation}
    GlobalAttention(\mathbf{H}^{O},\mathbf{C})={\mathrm{softmax}}(\mathbf{H}^{O}\mathbf{C}^{\mathsf{T}})\mathbf{C}
\end{equation}

Note that if we reject all statistical information (i.e., $\epsilon =0$),Eqn.(11) will become self-attention\cite{12} as $ H^O = C$.

After passing through fully-connected layers and a softmax layer, feature vectors are mapped to the label space for label prediction and loss calculation. To maximize the probability of the correct label $Y_{True}$, we deploy an optimizer tominimize cross-entropy loss $L$.

\begin{equation}
    L=\mathrm{CrossEntropy}(Y_{\mathrm{True}},Y_{\mathrm{Pred.}})
\end{equation}

\section{Experimental Setup}

\subsection{Dataset and Hyper-parameters}

We evaluate our model ASGNet on seven public datasets. The statistics of these seven datasets are listed in Table \ref{tab2}. 


\begin{table}[htbp]
  \centering
  \caption{The statistics of the seven datasets}
    \begin{tabular}{p{2cm}p{2cm}p{2cm}p{3cm}p{2cm}}
    \hline
    Datasets  & Size  & \#of logs  & \#of anomalies  & \#of anomalies types \\
    \hline
    BGL &  1.207GB    & 4,747,963    & 949,024  & 5  \\
    BGP  & 1.04GB  & 11,428,282    & 1,276,742 & 11\\
    Tbird &  27.367GB    & 211,212,192    & 43,087,287  & 7  \\
    Spirit  & 30.289GB   & 272,298,969    & 78,360,273 & 8\\
    HDFS &  1.58GB    & 11,175,629    & 362,793  & 6  \\
    Liberty  & 29.5GB   & 266,991,013    & 191,839,098 & 17\\
    Zookeeper &  10.4M    & 74,380   & 49,124  & 10  \\
    \hline
    \end{tabular}%
  \label{tab2}%
\end{table}%

\subsection{Training and hyperparameters}
We fix all the hyper-parameters applied to our model. Specifically, We use the basic version of RoBERTa as the pre-trained embeddings in our experiments. The $\epsilon$ set to 0.2, which empirically shows the best performance.  The algorithm we choose for optimization  is Adam Optimizer with the first momentum coefficient $\beta_{1}$= 0.9 and the second momentum coefficient $\beta_{2}$=0.999. We use the best parameters on development sets and evaluate the performance of our model on test sets.

\section{Experimental Results}

In this section, we elaborate the experimental setup and analyze the experimental results, aiming to answer:

\textbf{RQ1:} Can ASGNet achieve better log-based anomaly diagnosis performance than the state-of-the-art methods for log-based anomaly diagnosis task?

\textbf{RQ2:} How do the key model components and information types used in ASGNet contribute to the overall performance?

\textbf{RQ3:} How does the size of these parameters, specifically, hidden state dimension of global attention and the gate function $\epsilon$, affect the performance of the entire model?

\subsection{Model Comparisons (RQ1)}

To analyze the effectiveness of our model, we take some of the most advanced methods as baselines on the above-mentioned seven datasets, to validate the performance of our ASGNet model. The results are demonstrated as follows.

\begin{figure}[htbp]
\begin{center}
    \includegraphics[width=\textwidth]{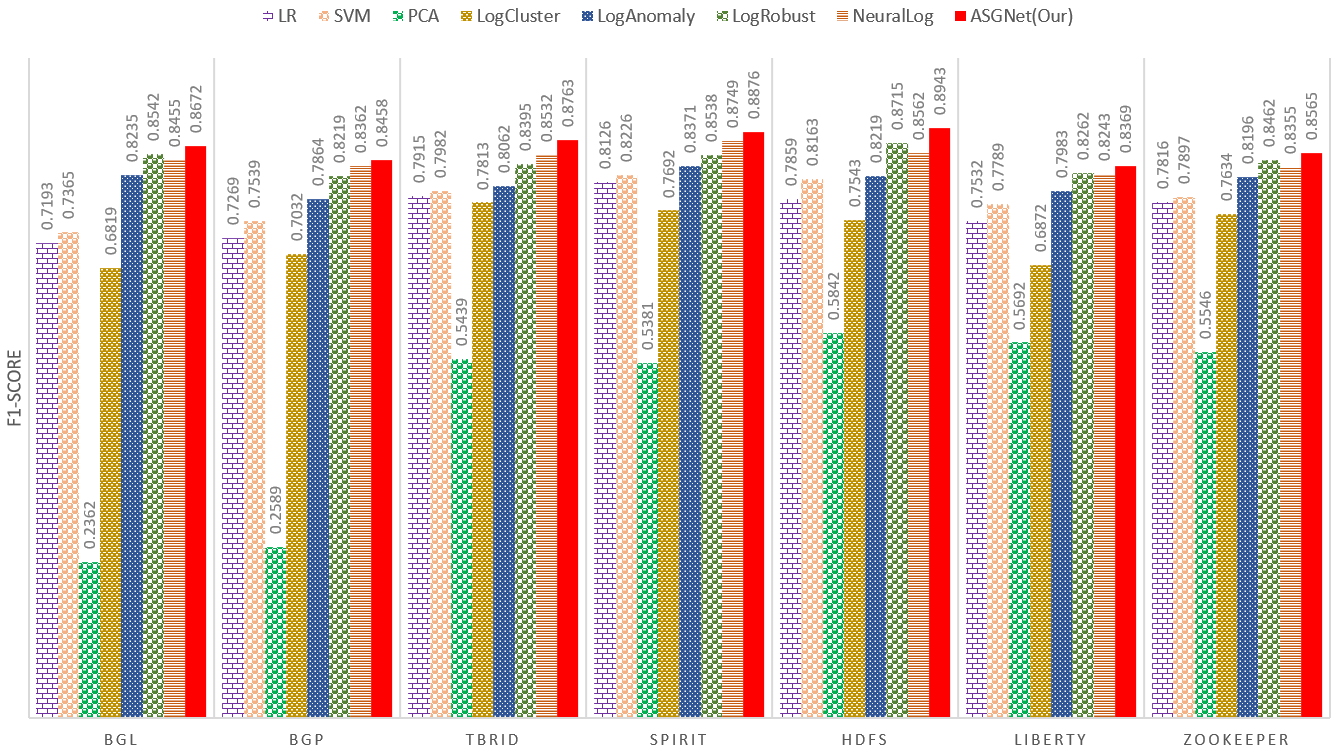}
\end{center}
\caption{Overall perform of our proposed ASGNet model.}
\label{f3}
\end{figure}

As shown in Figure \ref{f3}, \textbf{NeuralLog} \cite{14} , \textbf{LogRobust} \cite{9} and \textbf{LogAnomaly} \cite{13} are strongest baseline models of the Log-Based Anomaly Detection task. The experiment results show that our model ASGNet achieves best performance on seven datasets. For instance, 
on the BGL dataset, our model outperforms baseline model \textbf{NeuralLog} \cite{14} by 2.17\% in F1-Score (p $<$ 0.05 on student t-test). \textbf{LogRobust} \cite{9}  by 1.3\% in F1-Score (p $<$ 0.05 on student t-test). \textbf{LogAnomaly} \cite{13} by 4.37\% in F1-Score (p $<$ 0.05 on student t-test). It also showed good performance compared to the comparison method on the other six datasets.

The reason is our proposed model takes into account not only the semantic features behind the log execution logic but also the statistical features of the log text in the log exception diagnosis task. In order to better fit the two features, we propose well-designed threshold mechanism which effectively selects the statistical features to consolidate the semantic features, instead of using all statistical features. The threshold mechanism introduces the information flow into the classifier based on the confidence of the semantic feature in the decision, which is conducive to training a robust classifier and can solve the overfitting problem caused by the use of statistical features.Therefore the final experimental results demonstrate that our method achieves remarkable performance in the log-based anomaly diagnosis task.

\subsection{Ablation Study (RQ2)} 
To thoroughly figure out the effect of each key model component, we carry out a series of ablation study to decompose the whole model into three derived models. 

\begin{figure*}[htbp]
  \centering
  \includegraphics[width=\textwidth]{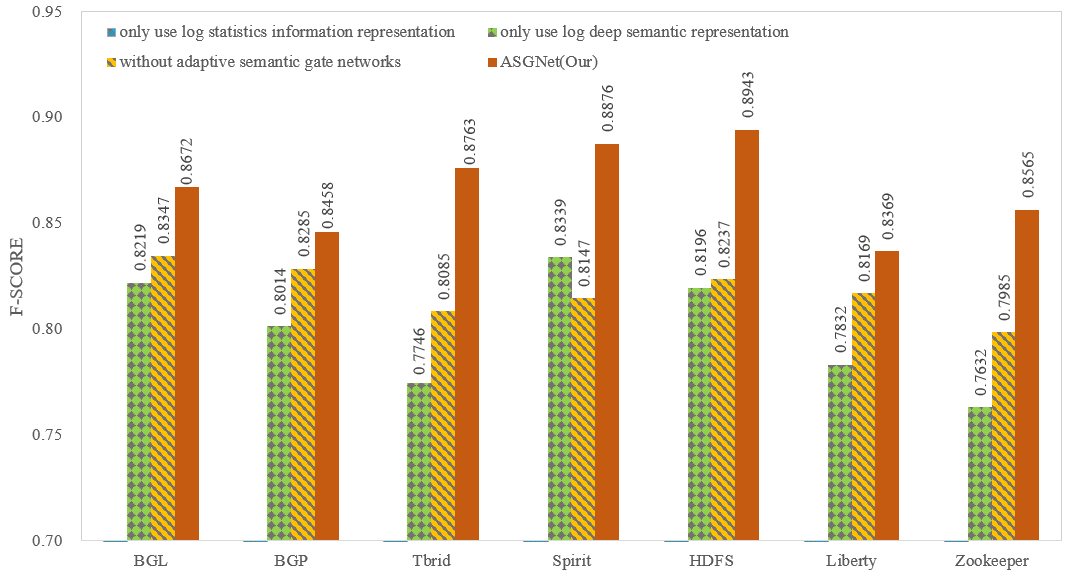}
  \caption{Ablation study on seven datasets}
  \label{f4}
\end{figure*}

\textbf{Model (1): only use log statistics information representation} -- The entire model only uses the statistics information Representation of the log sequence.

\textbf{Model (2): only use log deep semantic representation} -- The entire model only uses the log deep semantic representation of the log sequence.

\textbf{Model (3): without adaptive semantic gate networks} -- The entire model excludes the adaptive semantic gate networks.

As shown in Figure \ref{f4}, we can see that in our model, the log deep semantic representation module contributes more to task log-based anomaly diagnosis than the log statistics information representation module, mainly because the semantic information of the text of the logs can better express the deep logical semantic information of the logs, while the statistical features can only express the high and low frequency distribution of words, so log deep semantic representation module shows more advantages. In addition, it can be seen that the adaptive semantic gate module is indispensable in the whole component of our model, and removing the adaptive semantic gate module will affect the overall performance of the text proposed model.

\subsection{Parameter Sensitivity (RQ3)}

\begin{figure}[htbp]
  \centering
  \includegraphics[width=\linewidth]{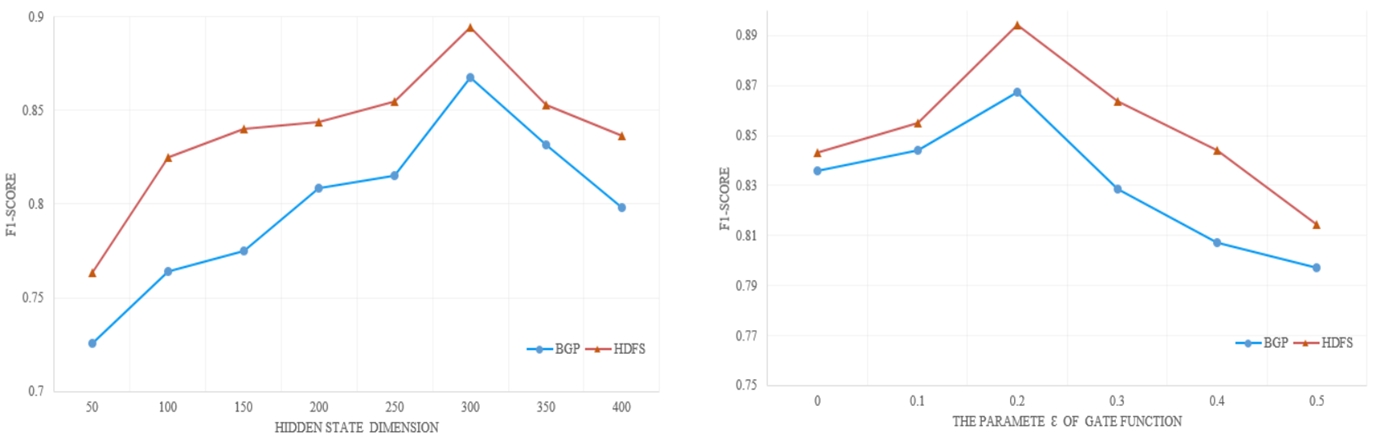}
  \caption{Performance of model ASGNet influenced by different hidden state dimension of global attention and the gate function $\epsilon$}
  \label{f5}
\end{figure}

As shown in Figure \ref{f5}, firstly, we can see that on two datasets the f1-score of our model shows an upward trend when the dimension size is less than 300, especially achieving highest when the dimension size is exactly 300, which indicates that a large dimension size could contribute to model performance. However, when the dimension size is larger than 300, the F1-Score of the model drops on both HDFS and BGL Dataset,  possibly due to insufficient training data. 

Secondly, we compare the performance of our model using the gate function $\epsilon$ on two datasets. As illustrated in Figure \ref{f5}, our model achieves best f1-score with the $\epsilon = 0.2$ thus demonstrating that not all statistical features are useful for the model. 

\section{Conclusion}
Anomaly detection and diagnosis play an important role in the event management of large-scale systems, which aims to detect abnormal behavior of the system in time. Timely anomaly detection enables system developers (or engineers) to pinpoint problems the first time and resolve them immediately, thereby reducing system downtime. In this paper we design an Adaptive Semantic Gate Networks (ASGNet) which consists of three parts, log statistics information representation(V-Net),log deep semantic representation(S-Net) and adaptive semantic threshold mechanism(G-Net). Specifically, V-Net leverages an unsupervised autoencoder to learn a global representation of each statistical feature vector. S-Net extracts latent semantic representations from text input by pre-trained RoBERTa. G-Net aligns information from two sources, then adjusts the information flow.The final experimental results demonstrate that our method achieves remarkable performance in the log-based anomaly diagnosis task.

\section{Acknowledgement}
This  work  is  partially  supported  by  E0HG043104.

%
%
%
%
%
\bibliographystyle{unsrt}
\bibliography{mybibliography}
%





\end{CJK}
\end{document}